# High Stellar FUV/NUV Ratio and Oxygen Contents in the Atmospheres of Potentially Habitable Planets

Feng Tian[1,2], Kevin France[3], Jeffrey L. Linsky[4], Pablo J.D. Mauas[5], Mariela C. Vieytes[5]

1. National Astronomical Observatories of China, Beijing, China
2. Center for Earth System Sciences, Tsinghua University, Beijing, China 100084
3. Center for Astrophysics and Space Astronomy, University of Colorado, 389 UCB, Boulder, CO 80309, USA
4. JILA, University of Colorado and NIST, 440 UCB, Boulder, CO 80309-0440, USA
5. Instituto de Astronomía y Física del Espacio (CONICET-UBA), CC. 67, Suc. 28 (1428), Buenos Aires, Argentina

**Abstract: Recent observations of several planet-hosting M dwarfs show that most have FUV/NUV flux ratios 1000 times greater than that of the Sun. Here we show that the atmospheric oxygen contents ($O_2$ and $O_3$) of potentially habitable planets in this type of UV environment could be 2~3 orders of magnitude greater than those of their counterparts around Sun-like stars as a result of decreased photolysis of $O_3$, $H_2O_2$, and $HO_2$. Thus detectable levels of atmospheric oxygen, in combination with the existence of $H_2O$ and $CO_2$, may not be the most promising biosignatures on planets around stars with high FUV/NUV ratios such as the observed M dwarfs.**

## 1. Introduction

Because of current observation constraints, M dwarfs are the most promising targets in the search for habitable exoplanets and life on such planets in the near future (Anglada-Escudé et al. 2013). The Earth's atmosphere remained mostly anoxic until 2.2~2.4 billion years ago (GOE), and the rise of $O_2$ in the Earth’s atmosphere after that time is widely accepted to be the result of the evolution of oxygenic photosynthesis (Kasting 1993). Today the oxygen abundances in the Earth's atmosphere are maintained orders of magnitude out of chemical equilibrium only by biological activity (Des Marais et al. 2002, Kaltenegger et al. 2007, Segura et al. 2005, Léger et al. 2011). Observationally, atmospheric $O_2$ with a mixing ratio exceeding 0.2% could be detected at 0.76 micron by a future optical/near-IR (TPF-C type) mission (Kaltenegger et al. 2007), and $O_3$ with a column density $>10^{17}$ $cm^{-2}$ could be detected at its 9.6 micron band by a future thermal IR (TPF-I or Darwin-like) mission (Kaltenegger et al. 2007, Segura et al. 2003).

Many previous papers (Walker 1977, Kasting et al. 1979, Kasting et al. 1984, Kasting 1990, 1993, 1995, 1997, Rosenqvist and Chassefiere 1995, Segura et al. 2003, 2005, 2007) have been published regarding atmospheric oxygen as a biosignature. Selsis et al. (2002) pointed out that in an atmosphere with high $CO_2$ content (0.2 bar to several bars), abiotic atmospheric oxygen could build up in some cases but the high $CO_2$ contents in such atmospheres mask the $O_3$ IR features completely. Thus if the future observations of exoplanetary atmosphere show high abundances of $CO_2$, abiotic oxygen build-up is possible and the source of atmospheric oxygen, if detected, should be carefully examined. A more recent paper (Hu et al. 2012) supports the Selsis et al. results about oxygen abiotic buildup in a dense $CO_2$ atmosphere.

The existence of detectable levels of atmospheric $O_2$ and/or $O_3$ with $H_2O$ in a planet's atmosphere with limited concentration of $CO_2$ (lower than 0.2 bar) is considered by some researchers as one of the most promising biosignature for habitable planets (Léger et al. 1996, Des Marais et al. 2002, Selsis et al. 2002, Segura et al. 2003, 2005, 2007, Kaltenegger et al. 2007, Léger et al. 2011). However, atmospheric oxygen contents have not been studied using realistic UV spectra of planet-hosting M dwarfs. Most previous observations of M dwarfs are limited to wavelengths longer than 1850 Å (Buccino et al. 2007, Walkowicz et al. 2008), which do not include the FUV bandpass (1170-1750 Å) or the bright Lyman-α emission line (Linsky et al. 2013). IUE did make observations of several inactive M dwarfs, but its sensitivity and spectral resolution were insufficient to make significant detections of FUV emission lines. The Ly-α fluxes are completely dominated by geocoronal airglow in these observations. The Living with a Red Dwarf program did obtain integrated emission line fluxes (C III 977Å and O VI 1032Å) of a few M dwarfs (Engle et al. 2009). However, no Lyman-α (1216 Å) data were available except for 4 active M dwarfs (Wood et al. 2005). Thus previously observed UV spectra of M dwarfs are insufficient for computing atmospheric $O_2$ and $O_3$ abundances in planetary atmospheres.

New FUV and NUV observations of 6 planet-hosting M dwarfs by the Hubble Space Telescope show that 5 of them are strong emitters at FUV wavelengths but weak emitters at NUV wavelengths and one of them GJ1214 remains undetermined because of large signal-to-noise ratio (Frace et al. 2012, 2013),. For GJ876, the time-averaged integrated energy flux in Lyman-α on a planet at 0.21 AU (the outer part of the habitable zone of M dwarf GJ876) is 2 times that received by the Earth; but at NUV wavelengths, the Earth receives more than 2000 times the flux incident on the GJ876 planet. Even for GJ1214, which displays only weak (possibly intermittent) FUV and NUV radiation, its FUV/NUV ratio appears to be at least 500~600 times that of the Sun, albeit with large uncertainties (France et al. 2013).

FUV photons, originating in stellar chromospheres and transition regions, dissociate $CO_2$ to produce atomic oxygen. 3-body recombination of atomic oxygen can form $O_2$; 3-body recombination reaction between atomic oxygen and $O_2$ can form $O_3$. From this perspective FUV photons are net sources of atmospheric oxygen. NUV photons (1750-3200 Å), originating in both stellar photospheres (the continuum part) and chromospheres (the emission lines), dissociate $O_3$, $H_2O_2$ and $HO_2$ and other species but not $CO_2$. Thus high FUV/NUV photon flux ratio could have significant implications for the atmospheric composition of planets. In this work we study this effect using a 1-D photochemical model.

## 2. Model Descriptions

We apply the observed FUV and NUV spectra of GJ876 to a 1-D photochemical model (Tian et al. 2010, 2011) to study the photochemistry of a hypothetical abiotic Earth-mass planet near the outer edge of the habitable zone of GJ876. The FUV/NUV ratios of other observed weakly active M dwarfs are similar to that of GJ876. Thus we do not expect our conclusions to change significantly when using the UV spectra of other observed M dwarfs. For comparison, we also model an abiotic Earth using the current solar FUV and NUV data. Fig. 1 shows the FUV and NUV energy fluxes incident on the top of the atmospheres, as well as the absorption cross sections of 6 species.

Our model contains 52 long-lived species, 27 short-lived species, and 434 chemical reactions. The atmospheric eddy diffusivity and density profiles are identical for the GJ876 and the lifeless Earth models. To discuss the influence of different temperature profiles on the results, we use three temperature profiles corresponding to surface temperatures of 275, 288, and 298 K. In the remainder of this paper, the results using 288 K surface temperature will be discussed when the temperature is not specified. The details of the species, boundary conditions, reactions, and profiles are included in the supplemental materials.

For both planetary atmospheres, we use a fixed volcanic outgassing $H_2$ flux of $2.5\times10^{12}$ moles $H_2$/yr, or $10^{10}$ $H_2$ molecules cm$^{-2}$ s$^{-1}$, a value appropriate for the modern Earth (Tian et al. 2005), and set the $H_2$ surface deposition velocity to zero, reflecting the lack of consumption by organisms on the two lifeless planets. Hydrogen escape to space is assumed to be diffusion-limited (Walker 1977). Lightning is ignored for simplicity. To fully account for the dissociation of $CO_2$ and the transport effect of ozone, we include both in the model as long-lived species. Rainout and surface deposition of most oxidizing and reducing species are treated the same way as in previous works (Tian et al. 2010, 2011). For CO, we applied a fixed deposition velocity of $10^{-6}$ cm/s at the surface, suitable for an ocean in which CO reacts with the oxidizing species deposited from the atmosphere. The $O_2$ deposition velocity is set to equal to that of CO, consistent with their similar diffusivities and Henry's law constants.

In this work we ignore the burial of organics, which should help to increase the atmospheric oxygen content. In addition to $H_2$ outgassing, ferrous iron released from the interior of the planet is a source of reducing power to the atmosphere-ocean system. For the modern Earth, the global ferrous iron flux from hydrothermal systems (Kump and Holland 1992) is $2.5\times10^{10}$ moles Fe/yr, or $1.6\times10^{8}$ cm$^{-2}$ s$^{-1}$, and the upper limit for the iron deposition rate on early Earth is $3\times10^{12}$ moles of Fe/yr, or $1.2\times10^{10}$ cm$^{-2}$ s$^{-1}$, derived from the Hamersley BIF (Holland 2006). Atmospheric to ocean $O_2$ fluxes of $2.5\times10^{7}$ cm$^{-2}$ s$^{-1}$ and $2\times10^{9}$ cm$^{-2}$ s$^{-1}$, respectively, are needed to oxidize these ferrous iron inputs (note that 1 mole of $O_2$ consumes 2 moles of $H_2$ via the reaction $2\ H_2 + O_2 \rightarrow 2H_2O$, while 3 moles of ferrous iron generates 1 mole of $H_2$ when oxidized to magnetite: $3\ FeO + H_2O \rightarrow Fe_3O_4$). These fluxes are much smaller than the model predicts for $O_2$ deposition fluxes in the GJ876 UV case.

Most previous models only considered the redox balance of the atmosphere (Segura et al. 2003, 2005, 2007, Tian et al. 2010, 2011). The oceanic redox balance was considered for an Archean Earth with an active marine biosphere (Karecha et al. 2005). In the present work, we consider the redox balances of both the atmosphere and the ocean. As shown in Fig. 2, the volcanic outgassing of hydrogen ($\Phi_{Volc}$) and the rainout of oxidizing species from the atmosphere ($\Phi_{Oxi}$) are the sources of the reducing power of the atmosphere, while the escape of hydrogen ($\Phi_{Esc}$) and the rainout of reducing species ($\Phi_{Red}$) are the sources of the atmosphere's oxidizng power.

In addition to the rainout of reducing and oxidizing species from the atmosphere to the ocean, our model includes ferrous iron injection from hydrothermal systems into the ocean and the release of reducing gases back to the atmosphere. The rainout of reducing species from the atmosphere and the ferrous iron injection are sources of the ocean's reducing power, and the rainout of oxidizing species is a sink of the ocean's reducing power. On a geological timescale, the redox balance in the ocean cannot be satisfied automatically by these terms. Thus in order to reach long term ocean redox balance, an upward flux ($\Phi_{Ocean}$) of reducing gases (in the form of $H_2$) is added at the lower boundary of the atmosphere model with the value equal to the difference between the ferrous iron flux and the rainout fluxes. The same ocean release flux is included in the atmospheric redox budget calculations.

We carried out simulations with a 1 bar surface pressure. Atmospheric $CO_2$ concentration is between 1 and 10%. More details are provided in the Supplemental Materials.

## 3. Results and Discussions

High $O_2$ and $O_3$ contents are obtained for the UV radiation input from GJ876 but not for the Sun. Fig. 3 shows the typical profiles of $O_2$, $O_3$, CO, and $CO_2$ (fixed to 5% at the surface) in the two cases. In the solar UV case, the calculated $O_2$ mixing ratio decreases from $10^{-3}$ at 60 km altitude to $10^{-17}$ at the surface, and the $O_2$ vertical column density is $3.7\times10^{19}$ $cm^{-2}$, similar to the results in previous papers (Segura et al. 2007). In contrast, the $O_2$ mixing ratio in the GJ876 UV case remains ~0.4% everywhere below 70 km, and the $O_2$ vertical column density is ~$8\times10^{22}$ $cm^{-2}$. The calculated $O_3$ vertical column density is $1.5\times10^{15}$ $cm^{-2}$ in the solar UV case and $3.6\times10^{17}$ $cm^{-2}$ (240 times larger) in the GJ876 UV case. Using the UV spectrum of GJ876 clearly increases atmospheric oxygen content.

In order to understand the mechanisms, in the following table we compare the photolysis J values (the rate coefficient at which photolysis occurs for a certain species) at the top of the atmosphere (TOA) of several key species in the solar and GJ876 UV models. It is clear that the main effects of using the GJ876 UV spectrum is to decrease the photolysis of $O_3$, $H_2O_2$, and $HO_2$.

| J ($s^{-1}$) | $J_{CO2}$ | $J_{O2}$ | $J_{O3}$ | $J_{H2O}$ | $J_{H2O2}$ | $J_{HO2}$ |
|---|---|---|---|---|---|---|
| Solar | 1.5E-7 | 2.6E-6 | 6.8E-3 | 7.8E-6 | 7.8E-5 | 5.0E-4 |
| GJ876 | 1.9E-7 | 2.5E-6 | 4.3E-5 | 8.4E-6 | 7.3E-7 | 1.3E-6 |

To further test the effects of different photolysis efficiencies of $O_3$, $H_2O_2$, and $HO_2$ on our results, we carried out sensitivity tests in the 5% $CO_2$ + 288 K surface temperature case using the GJ876 UV spectrum but artificially increasing the photolysis J values of $O_3$, $H_2O_2$, and $HO_2$ by factors of 100, 100, and 500, respectively. When we increase only $J_{O3}$, the $O_3$ content decreases 30 times but the $O_2$ content decreases only 50%. When we increase $J_{H2O2}$ and $J_{HO2}$ at the same time, $O_3$ decreases by 3 orders of magnitude and $O_2$ decreases by 30 times. On the other hand, when we inhibit the photolysis of $O_3$, $H_2O_2$, and $HO_2$ in the solar UV case, high oxygen contents similar to those in the GJ876 UV cases are obtained.

These sensitivity tests suggest that although enhanced $O_3$ photolysis decreases $O_3$ in the GJ876 UV model, it does not influence the $O_2$ content significantly. Slow $H_2O_2$ and $HO_2$ photolysis is the fundamental reason for the high oxygen content in such an atmosphere. As shown in the lower panel of Fig. 1, while the cross sections of $CO_2$ and $O_2$ are significant only in the FUV region, the cross sections of $O_3$, $H_2O_2$, and $HO_2$ are significant in the NUV, where the solar and GJ876 UV spectra differ most.

Both $H_2O_2$ and $HO_2$ photolysis are sources of OH radicals. Selsis et al. (2002) investigated the following catalytic oxygen loss mechanism involving $HO_2$:

| | | | | | |
|---|---|---|---|---|---|
| (1) | CO | + OH | → | $CO_2$ | + H |
| (2) | H | + $O_2$ + M | → | $HO_2$ | + M |
| (3) | O | + $HO_2$ | → | $O_2$ + OH | |

-------------------------------------------------------------

**CO + O → $CO_2$**

For $H_2O_2$, similar but less efficient catalytic oxygen loss mechanism can be written:

| | | | | | |
|---|---|---|---|---|---|
| (1) | CO | + OH | → | $CO_2$ | + H |
| (2) | H | + $O_2$ + M | → | $HO_2$ | + M |
| (4) | O | + $H_2O_2$ | → | $HO_2$ | + OH |
| (5) | $HO_2$ | + $HO_2$ | → | $H_2O_2$ | + $O_2$ |

-------------------------------------------------------------

**CO + O → CO2**

Thus slower photolysis of $HO_2$ and $H_2O_2$ decreases the availability of OH, which in turn slows down these catalytic oxygen loss mechanisms. All previous works (Selsis et al. 2002, Segura et al. 2003, 2005, 2007) use either increased or decreased total UV fluxes but did not consider a much higher FUV/NUV ratio than that of the Sun. The effect of the combination of rapid $CO_2$ photolysis with slow $H_2O_2$ and $HO_2$ photolysis on atmospheric oxygen contents were not studied.

Fig. 4 shows column densities of atmospheric $O_2$ and $O_3$ (multiplied by $10^5$) in the GJ876 UV model as functions of CO2 concentrations, regulated by surface and atmospheric temperature. We used three temperature profiles, corresponding to surface temperatures of 275 (colder climate), 288, and 298 K (warmer climate), respectively, as shown in the supplemental materials. The water vapor content in the warmer climate case is orders of magnitude greater than that in the colder climate case (shown in the supplemental materials). The upper solid and dashed curves are for the colder climate and the lower ones are for the warmer climate. Because photolysis of $H_2O$ produces H and OH radicals which remove $O_2$, it is more difficult to accumulate oxygen in a warmer and wetter atmosphere (thus more water vapor). But in all cases the oxygen contents increase with $CO_2$ concentrations because $H_2O$ photolysis could be reduced by an increased content of atmospheric $CO_2$.

Our hypothetical GJ876 planet is placed in the outer part of the habitable zone, while planets in the inner part of the habitable zone receive a higher dosage of UV photons and should have faster photochemistry. Doubling the input FUV and NUV fluxes in the GJ876 UV case for 5% $CO_2$ + 288 K surface temperature increases the $O_2$ and $O_3$ column densities to $1.3\times10^{23}$ $cm^{-2}$ and $5.7\times10^{17}$ $cm^{-2}$ ,respectively, both of which are factors of 60% larger.

On the geological timescale, a planet's atmospheric $CO_2$ content is linked to the planet's climate (warmer climate favors conversion of silicates to carbonates and the reduction of atmospheric $CO_2$, colder climate favors an increase of atmospheric $CO_2$; Walker 1977). Therefore, planets in the inner part of a habitable zone should have less atmospheric $CO_2$ than those in the outer part. Thus planets in the outer part of the habitable zones of M dwarfs may accumulate higher amount of $O_2$ and $O_3$ in their atmospheres than planets in the inner habitable zones. Whether or not this trend is reversed by the different UV dosage reaching the planets depends on the UV brightness and particular spectral shape of the central star.

Because we do not have the capability for calculating a transmission or a thermal emission spectrum, we rely on the results of previous published works, which show that an $O_2$ concentration of 0.2% and $O_3$ column density between $10^{17}$ and $10^{18}$ $cm^{-2}$ in exoplanetary atmospheres with 0.01 bar $CO_2$, similar to that on early Earth at 2 Ga, are detectable by future missions to search for life on exoplanets (Kaltaegger et al. 2007, Segura et al. 2003), provided that the atmospheric $CO_2$ content is low (Selsis et al. 2002). The calculated $O_2$ contents in the GJ876 UV model exceed 0.2% by a safe margin. Therefore, positive detections of $O_2$, $H_2O$ and $CO_2$ in the atmospheres of habitable planets around stars with high FUV/NUV ratios do not necessarily imply the existence of life. Detectable levels of $O_2$ but not $O_3$ were obtained in Selsis et al. (2002) for both dry and humid atmospheres without Earth-like rainout and outgassing rates. Our work focus on Earth-like conditions. Selsis et al. (2002) also obtained high $O_3$ contents in $CO_2$-dominant atmospheres. By comparison, the calculated $O_3$ contents in our work are for $N_2$-dominant atmospheres and thus could be detectable. Future spectral analysis is needed to determine whether the combination of $O_3$, $H_2O$ and $CO_2$ is a true biosignature, as proposed by Selsis et al. (2002), around stars with high FUV/NUV ratios.

The readers are cautioned that our calculations are appropriate for habitable planets with Earth-like environments. Not all exoplanets in their habitable zones have similar hydrological and outgassing activities as those of the Earth. Thus even if the abiotic oxygen contents for some particular set of parameters is low, this does not prove that a high oxygen content is a 100% reliable biosignature. Thorough studies of environmental parameter space such as those in Selsis et al. (2002) are needed to support this claim. On the other hand, to find a "false positive" result for only one exoplanet with plausible hydrological and outgassing activities, for which the Earth's rainout and outgassing rates are at least realistic choices, would be adequate to claim that such a biosignature is not 100% reliable. It is noted that the coupling between the visible-NIR irradiance of the star and the planet's atmospheric thermal profile might impact the results by strongly affecting the efficiency of the tropospheric cold trap and thus the amount of water vapor in the stratosphere.

Because high $CO_2$ content favors high abiotic $O_2$ and $O_3$ contents, quantitative determination of $CO_2$ in exoplanetary atmospheres through observations can provide insights concerning the origin of the observed oxygen. Our simulations also show that whenever high abiotic atmospheric $O_2$ and $O_3$ appear, high CO contents also appear in the atmosphere. Thus the determination of CO abundances in habitable exoplanetary atmospheres could provide constraints on the origin of observed oxygen. In this aspect, ground-based high resolution NIR spectroscopy could be useful (Kok et al. 2013). Because of the importance of FUV and NUV radiation in driving oxygen photochemistry, observations of the FUV and NUV spectra of planet-hosting stars are critical. A robust claim for the detection of biosignatures in exoplanetary atmospheres requires the combination of observations and photochemical models.

## 5. Conclusions

Until now, the combination of $O_2$ and/or $O_3$, $H_2O$, and $CO_2$ in the atmosphere of a rocky planet in the habitable zone has been considered the most promising biosignature, especially when considering Earth-like hydrological and outgassing activities. But previous models have not included realistic FUV/NUV flux ratios from planet-hosting M dwarfs. We find that the inclusion of realistic UV fluxes from such stars calls into question the reliability of this combination of molecules as a biosignature.

Recent observations of 6 planet-hosting M dwarfs unambiguously show that most have FUV/NUV ratios 3 orders of magnitude larger than that of the Sun. We show here that slow $H_2O2$ and $HO_2$ photolysis, driven by this unique combination of UV spectra, weakens the catalytic oxygen loss mechanisms in planetary atmospheres. As a result, the atmospheric $O_2$ content on a habitable but lifeless planet with Earth-like hydrological and outgassing activities could accumulate to levels detectable by future missions that search for life when the central star has a high FUV/NUV ratio similar to those of the recently observed planet-hosting M dwarfs. $O_3$ photolysis in such a UV environment is also reduced and the atmospheric $O_3$ content on a lifeless habitable planet with Earth-like hydrological and outgassing activities increases by 2~3 orders of magnitude in comparison with that in solar EUV environment. The detectability of $O_3$ in such a case requires further spectral analysis.

Despite the potential ambiguity regarding atmospheric oxygen as a reliable biosignature, the search for life on exoplanets could be helped by the combination of 1) a quantitative determination of planetary atmosphere composition ($H_2O$, $CO_2$, and CO in addition to oxygen), 2) medium-to-high resolution stellar UV spectrum, and 3) appropriate photochemistry modeling of the planet's atmosphere.

# Acknowledgement:

We thank R. Hu and an anonymous reviewer for their constructive reviews which significantly improved the quality of this paper. FT thanks A. Léger and E. Chassefiere for helpful communications regarding the current understanding of atmosphere oxygen as biosignature.

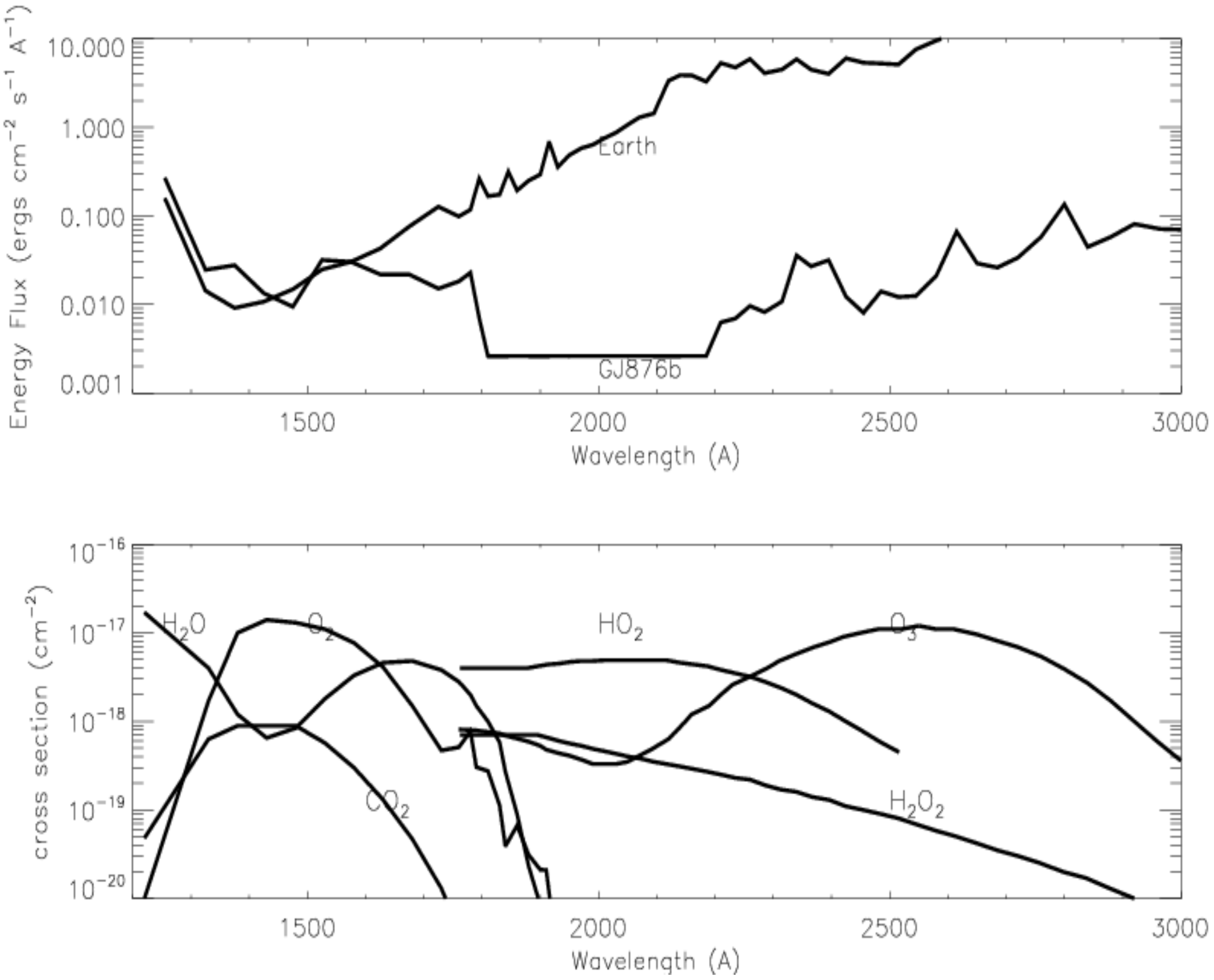

Fig. 1: Upper Panel: Time-averaged UV fluxes incident on a hypothetical lifeless Earth and a lifeless planet at 0.21 AU from GJ876 (France et al. 2013). The GJ876 NUV flux between 1800 and 2200 Å is below the observation limit and is set to this value, forming the flat line in the wavelength region in the GJ876b spectrum. Lower Panel: the photolysis cross sections of 6 species in the model. $H_2O$, $O_2$, and $CO_2$ photolysis is driven by FUV photons. $H_2O_2$, $HO_2$, and $O_3$ photolysis is driven by NUV photons.

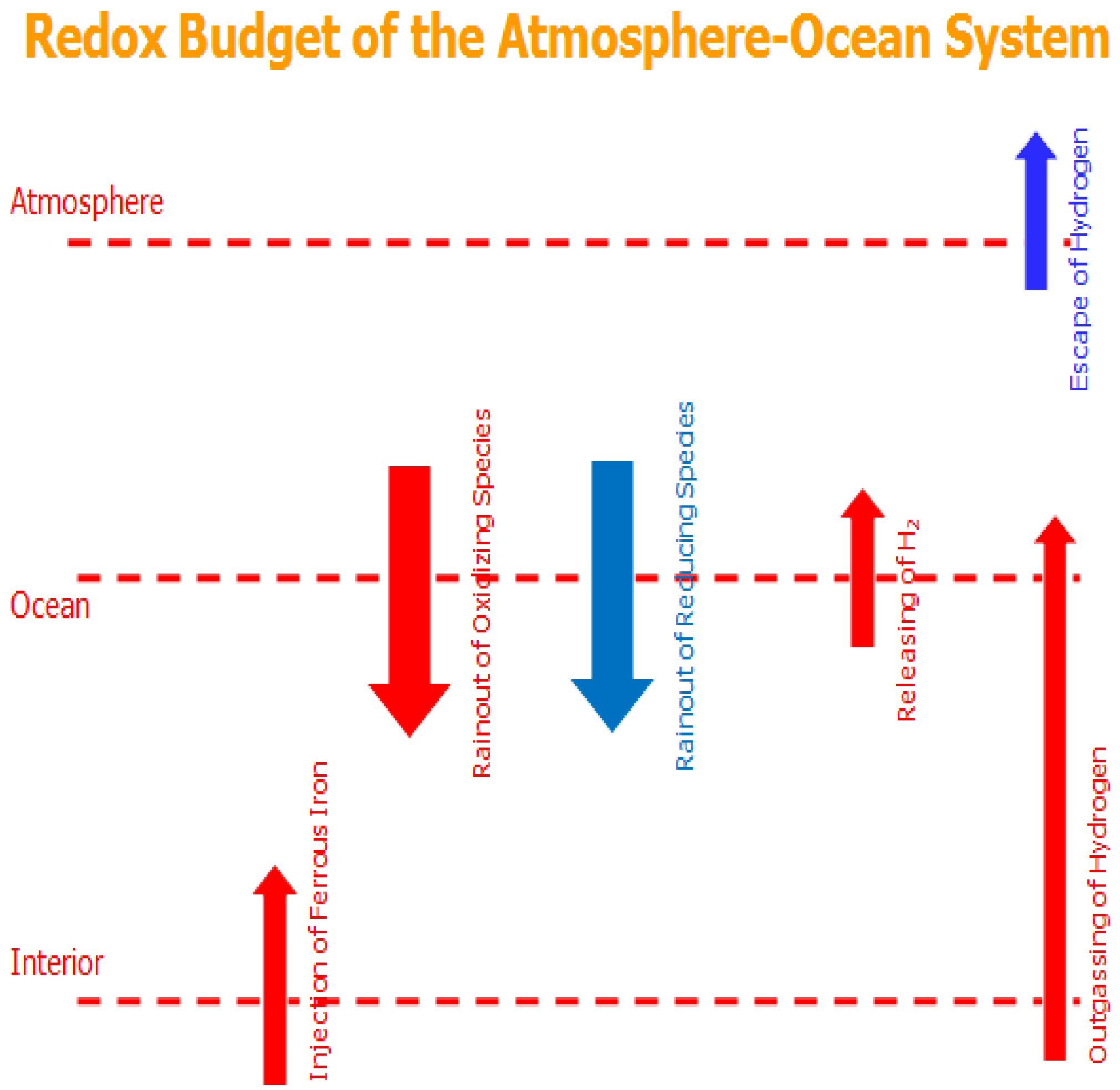

Fig.2: Redox Budget of the atmosphere-ocean system. In the atmosphere, the outgassing of hydrogen, rainout of oxidizing species, and release of hydrogen from the ocean are sources of reducing power (red color), while the escape of hydrogen and rainout of reducing species are sinks of the atmosphere's reducing power (blue color). In the ocean, ferrous iron from hydrothermal systems and rainout of reducing species from the atmosphere are sources of reducing power, while the rainout of oxidizing species and release of hydrogen from the ocean into the atmosphere are sinks of the ocean's reducing power. Burial of organic matters is ignored which could further increase atmospheric oxygen content if included in the redox budget. The ocean redox budget was not included in previous models, except that of Kharecha et al. 2005).

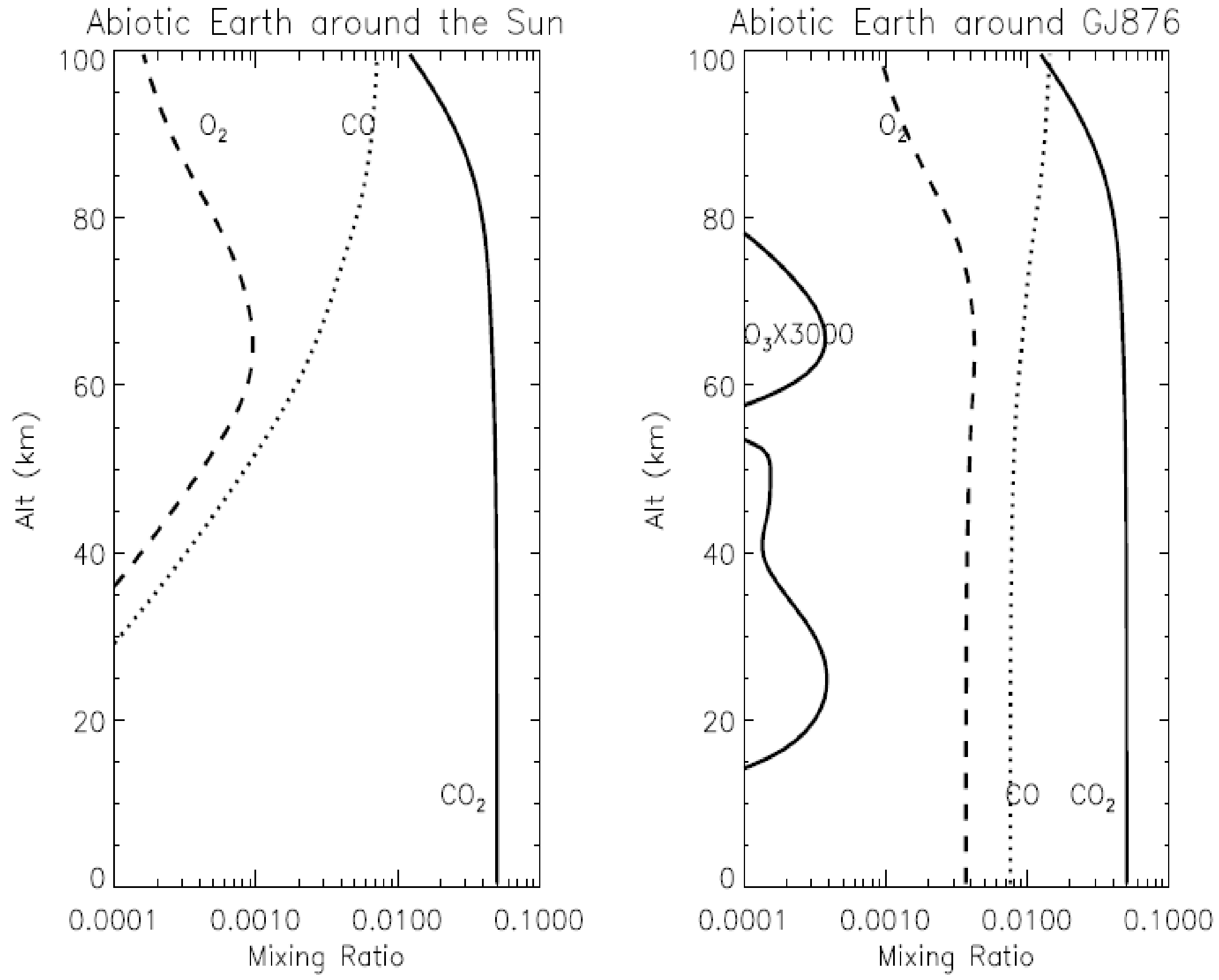

Fig. 3: Atmospheric profiles of $O_2$ (dashed), CO (dotted), $CO_2$ (solid curves on the right in both panels), and $O_3$ (solid curves on the left in the right panel) for abiotic Earth-mass planets around the Sun and GJ876. Atmospheric $CO_2$ levels are set to 5% at the surface on both planets. The $O_3$ mixing ratios in both panels are multiplied by a factor of 3000.

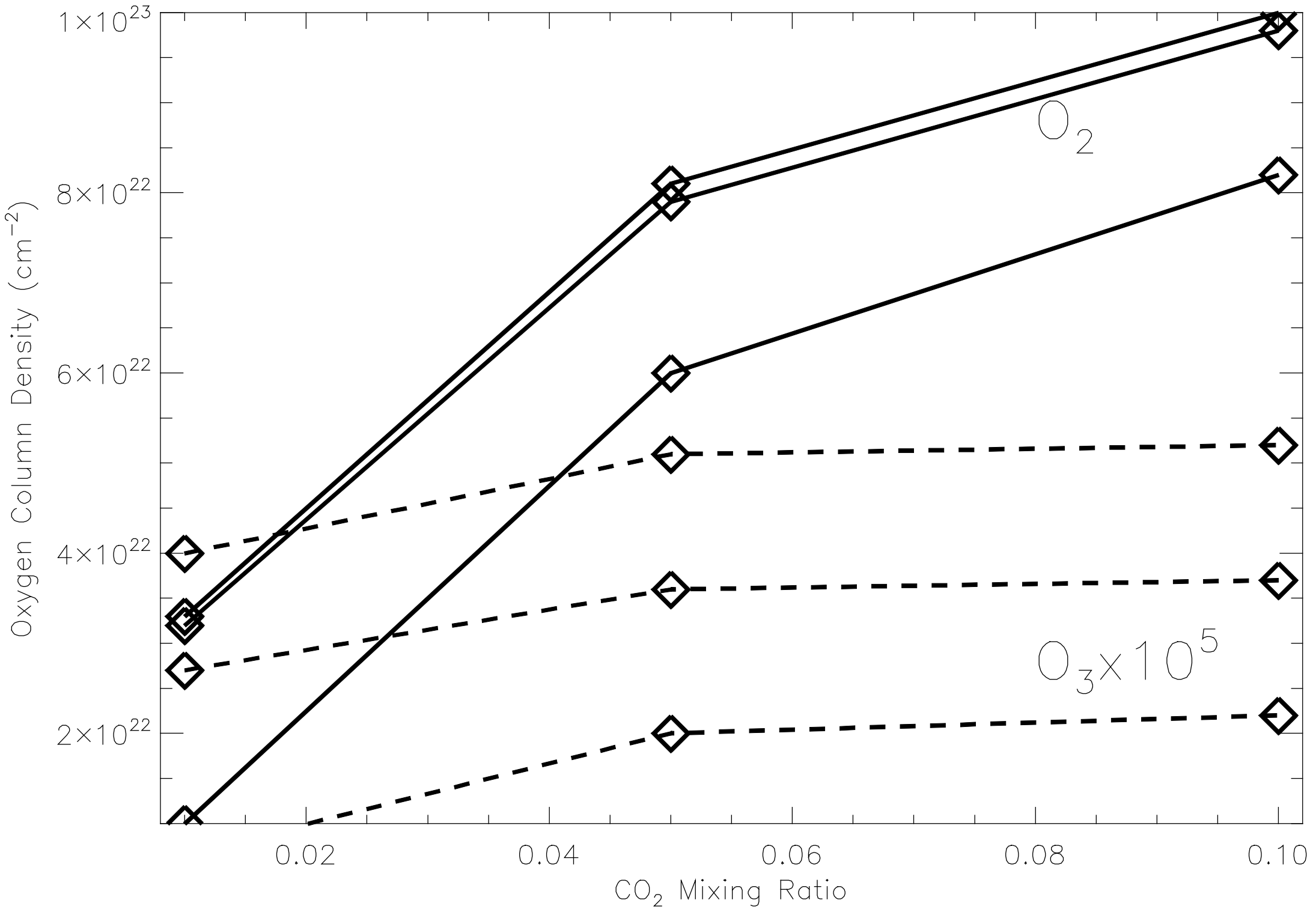

Fig. 4 Atmospheric oxygen contents as functions of $CO_2$ concentrations under different surface temperatures. The diamond symbols represent the specific cases modeled. The solid and dashed curves are for $O_2$ and $O_3$ (multiplied by $10^5$) column densities, respectively. The upper solid and dashed curves are for 275 K surface temperature cases, representing colder and drier climates, and the lower ones are for 298 K surface temperatures, representing warmer and wetter climates. The middle curves are for surface temperature of 288 K, the global average surface temperature of present Earth.